\input harvmac
\input amssym 

\def\p{\partial}

\def\rt{\rightarrow}

\def\go{g^{(0)}}
\def\gt{g^{(2)}}
\def\gf{g^{(4)}}
\def\gln{g^{(\rm ln)}}

\def\Ao{A^{(0)}}
\def\At{A^{(2)}}
\def\Th{\hat{T}}

\def\eps{\epsilon}

\def\aa{a^{(0)}}
\def\ab{a^{(1)}}

\def\Th{\hat{T}}

\def\azero{a^{(0)}}
\def\atwo{a^{(2)}}
\def\lamt{\tilde{\lambda}}
\def\cJ{{\cal J}}
\def\cO{{\cal O}}

\def\ep{\varepsilon}
\def\cJh{\hat{\cJ}}
\def\cTh{\hat{\cT}}
\def\cT{{\cal T}}

 
\lref\DHokerIJ{
  E.~D'Hoker and P.~Kraus,
  ``Magnetic Field Induced Quantum Criticality via new Asymptotically AdS$_5$
  Solutions,''
  arXiv:1006.2573 [hep-th].
}

\lref\deHaroXN{
  S.~de Haro, S.~N.~Solodukhin and K.~Skenderis,
  ``Holographic reconstruction of spacetime and renormalization in the  AdS/CFT
  correspondence,''
  Commun.\ Math.\ Phys.\  {\bf 217}, 595 (2001)
  [arXiv:hep-th/0002230].
}

\lref\KrausWN{
  P.~Kraus,
  ``Lectures on black holes and the AdS(3)/CFT(2) correspondence,''
  Lect.\ Notes Phys.\  {\bf 755}, 193 (2008)
  [arXiv:hep-th/0609074].
}

\lref\AnninosPM{
  D.~Anninos, G.~Compere, S.~de Buyl, S.~Detournay and M.~Guica,
  ``The Curious Case of Null Warped Space,''
  arXiv:1005.4072 [hep-th].
}

\lref\BalasubramanianRE{
  V.~Balasubramanian and P.~Kraus,
  ``A stress tensor for anti-de Sitter gravity,''
  Commun.\ Math.\ Phys.\  {\bf 208}, 413 (1999)
  [arXiv:hep-th/9902121].
}

\lref\DHokerMM{
  E.~D'Hoker and P.~Kraus,
  ``Magnetic Brane Solutions in AdS,''
  JHEP {\bf 0910}, 088 (2009)
  [arXiv:0908.3875 [hep-th]].
}

\lref\HeemskerkHK{
  I.~Heemskerk and J.~Polchinski,
  ``Holographic and Wilsonian Renormalization Groups,''
  arXiv:1010.1264 [hep-th].
}

\lref\FaulknerJY{
  T.~Faulkner, H.~Liu and M.~Rangamani,
  ``Integrating out geometry: Holographic Wilsonian RG and the membrane
  paradigm,''
  arXiv:1010.4036 [hep-th].
}

\lref\NickelPR{
  D.~Nickel and D.~T.~Son,
  ``Deconstructing holographic liquids,''
  arXiv:1009.3094 [hep-th].
  }
\lref\toappear{E.~D'Hoker and P.~Kraus,  to appear}
%

\lref\DHokerRZ{
  E.~D'Hoker and P.~Kraus,
  ``Holographic Metamagnetism, Quantum Criticality, and Crossover Behavior,''
  JHEP {\bf 1005}, 083 (2010)
  [arXiv:1003.1302 [hep-th]].
}

\lref\DHokerBC{
  E.~D'Hoker and P.~Kraus,
  ``Charged Magnetic Brane Solutions in AdS$_5$ and the fate of the third law of
  thermodynamics,''
  JHEP {\bf 1003}, 095 (2010)
  [arXiv:0911.4518 [hep-th]].
}

\lref\MaldacenaMW{
  J.~M.~Maldacena and C.~Nunez,
  ``Supergravity description of field theories on curved manifolds and a no  go
  theorem,''
  Int.\ J.\ Mod.\ Phys.\  A {\bf 16}, 822 (2001)
  [arXiv:hep-th/0007018].
}

\lref\BrownNW{
  J.~D.~Brown and M.~Henneaux,
  ``Central Charges in the Canonical Realization of Asymptotic Symmetries: An
  Example from Three-Dimensional Gravity,''
  Commun.\ Math.\ Phys.\  {\bf 104}, 207 (1986).
}

\lref\WittenUA{
  E.~Witten,
  ``Multi-trace operators, boundary conditions, and AdS/CFT correspondence,''
  arXiv:hep-th/0112258.
}

\lref\KrausNB{
  P.~Kraus and F.~Larsen,
  ``Partition functions and elliptic genera from supergravity,''
  JHEP {\bf 0701}, 002 (2007)
  [arXiv:hep-th/0607138].
}

\lref\KrausWN{
  P.~Kraus,
  ``Lectures on black holes and the AdS(3)/CFT(2) correspondence,''
  Lect.\ Notes Phys.\  {\bf 755}, 193 (2008)
  [arXiv:hep-th/0609074].
}

\lref\AlmuhairiRB{
  A.~Almuhairi,
  ``AdS$_3$ and AdS$_2$ Magnetic Brane Solutions,''
  arXiv:1011.1266 [hep-th].
}

\lref\FaulknerWJ{
  T.~Faulkner, H.~Liu, J.~McGreevy and D.~Vegh,
  ``Emergent quantum criticality, Fermi surfaces, and AdS2,''
  arXiv:0907.2694 [hep-th].
}

\lref\EdalatiBI{
  M.~Edalati, J.~I.~Jottar and R.~G.~Leigh,
  ``Transport Coefficients at Zero Temperature from Extremal Black Holes,''
  JHEP {\bf 1001}, 018 (2010)
  [arXiv:0910.0645 [hep-th]].
}

\lref\AstefaneseiDK{
  D.~Astefanesei, N.~Banerjee and S.~Dutta,
  ``Moduli and electromagnetic black brane holography,''
  arXiv:1008.3852 [hep-th].
}

\lref\BreitenlohnerJF{
  P.~Breitenlohner and D.~Z.~Freedman,
  Annals Phys.\  {\bf 144}, 249 (1982).
  }

\lref\weinberg{ S.~Weinberg, ``Gravitation and Cosmology: Principles and Applications of the General Theory of Relativity", John Wiley and Sons, Inc,  (1972)}

\lref\FG{
C. Fefferman and C. Robin Graham, �Conformal Invariants�, in {\sl Elie Cartan et les Math\'ematiques
d�aujourd�hui} (Ast\'erisque, 1985) 95.}

\lref\BalasubramanianRE{
  V.~Balasubramanian and P.~Kraus,
  ``A stress tensor for anti-de Sitter gravity,''
  Commun.\ Math.\ Phys.\  {\bf 208}, 413 (1999)
  [arXiv:hep-th/9902121].
}

\lref\deHaroXN{
  S.~de Haro, S.~N.~Solodukhin and K.~Skenderis,
  ``Holographic reconstruction of spacetime and renormalization in the  AdS/CFT
  correspondence,''
  Commun.\ Math.\ Phys.\  {\bf 217}, 595 (2001)
  [arXiv:hep-th/0002230].
}

\lref\GubserDE{
  S.~S.~Gubser, I.~R.~Klebanov and A.~W.~Peet,
  ``Entropy and Temperature of Black 3-Branes,''
  Phys.\ Rev.\  D {\bf 54}, 3915 (1996)
  [arXiv:hep-th/9602135].
}

\lref\BuchelGB{
  A.~Buchel and J.~T.~Liu,
  ``Gauged supergravity from type IIB string theory on Y(p,q) manifolds,''
  Nucl.\ Phys.\  B {\bf 771}, 93 (2007)
  [arXiv:hep-th/0608002].
}

\lref\GauntlettAI{
  J.~P.~Gauntlett, E.~O Colgain and O.~Varela,
  ``Properties of some conformal field theories with M-theory duals,''
  JHEP {\bf 0702}, 049 (2007)
  [arXiv:hep-th/0611219].
}

\lref\GauntlettMA{
  J.~P.~Gauntlett and O.~Varela,
  ``Consistent Kaluza-Klein Reductions for General Supersymmetric AdS
  Solutions,''
  Phys.\ Rev.\  D {\bf 76}, 126007 (2007)
  [arXiv:0707.2315 [hep-th]].
}

\lref\MarolfND{
  D.~Marolf and S.~F.~Ross,
  ``Boundary conditions and new dualities: Vector fields in AdS/CFT,''
  JHEP {\bf 0611}, 085 (2006)
  [arXiv:hep-th/0606113].
}

\lref\giamarchi{
T.~Giamarchi, 
``Quantum Physics in One Dimension", Oxford University Press}
%

\lref\HungQK{
  L.~Y.~Hung and A.~Sinha,
  ``Holographic quantum liquids in 1+1 dimensions,''
  JHEP {\bf 1001}, 114 (2010)
  [arXiv:0909.3526 [hep-th]].
}

\lref\MaityZZ{
  D.~Maity, S.~Sarkar, N.~Sircar, B.~Sathiapalan and R.~Shankar,
  ``Properties of CFTs dual to Charged BTZ black-hole,''
  Nucl.\ Phys.\  B {\bf 839}, 526 (2010)
  [arXiv:0909.4051 [hep-th]].
}

\lref\BalasubramanianSC{
  V.~Balasubramanian, I.~Garcia-Etxebarria, F.~Larsen and J.~Simon,
  ``Helical Luttinger Liquids and Three Dimensional Black Holes,''
  arXiv:1012.4363 [hep-th].
}


\Title{\vbox{\baselineskip12pt
}} {\vbox{\centerline {RG Flow of Magnetic Brane Correlators}}}
\centerline{ 
  Eric D'Hoker$^\dagger$\foot{dhoker@physics.ucla.edu}, Per
Kraus$^{\dagger}$\foot{pkraus@ucla.edu}, and Akhil Shah$^\ddagger$\foot{akhil137@gmail.com}}
\bigskip
\centerline{${}^\dagger$\it{Department of Physics and Astronomy}}
\centerline{${}$\it{University of California, Los Angeles, CA 90095,USA}}
\bigskip
\centerline{${}^\ddagger$\it{Department of Electrical Engineering}}
\centerline{${}$\it{University of California, Los Angeles, CA 90095,USA}}

\baselineskip15pt

\vskip .3in
 
\centerline{\bf Abstract}
 
The magnetic brane solution to five-dimensional Einstein-Maxwell-Chern-Simons theory
provides a holographic description of the RG flow from four-dimensional Yang-Mills theory 
in the presence of a constant magnetic field to a two-dimensional low energy CFT. 
We compute two-point correlators involving the U(1) current and the stress tensor,
and use their leading IR behavior to confirm the existence of a single chiral current 
algebra, and of left- and right-moving Virasoro algebras in the low energy CFT. The common
central charge of the Virasoro algebras is found to match the Brown-Henneaux formula,
while the level of the current algebra is related to the Chern-Simons coupling. 
The coordinate reparametrizations produced by the Virasoro algebras on the 
AdS$_3$ near-horizon geometry arise from physical non-pure gauge modes in the 
asymptotic AdS$_5$ region, thereby providing a concrete example for the emergence of 
IR symmetries.
Finally, we interpret the infinite series of sub-leading IR contributions to the correlators
in terms of certain double-trace interactions generated by the RG flow in the low energy CFT.

\Date{December, 2010}
\baselineskip13pt

\newsec{Introduction} 

A powerful feature of the AdS/CFT correspondence is that it provides a geometrical
description of RG flow, via the emergence of an extra holographic radial direction in the  
bulk spacetime. 
An RG flow between  UV and  IR CFTs is mapped to a bulk geometry that interpolates 
between one asymptotic AdS geometry at large radial coordinate $r$, and another at small $r$.   

The UV and IR CFTs may differ in their spacetime dimensionality. The IR dimensionality is
reduced compared to that in the UV if the low energy excitations are confined to propagate
in only some of the directions of the original space.   A simple example of this phenomenon
occurs for charged particles in a strong magnetic field.  Semi-classically, the particles undergo
circular motion at the cyclotron frequency around the magnetic flux lines, and the low
energy excitations correspond to the drift velocity parallel to the field.   The same
picture emerges from Landau level quantization, with the result that the effective dimensionality
of the system at low energies is reduced by two, compared to that at high energies.   
A D=3+1 CFT  in the presence of a magnetic field thus flows to a D=1+1 CFT at low 
energies\foot{Strictly speaking, the two spatial dimensions should be compact in order
for the IR CFT to have a finite total central charge, rather than a central charge per unit area.} 

A bulk geometry dual to such an RG flow was found in \DHokerMM.\foot{An earlier related example
is studied in \MaldacenaMW, and some generalizations appear in \AlmuhairiRB.  We also note
that correlators in RG flows to spacetimes with an AdS$_2$ factor have been studied recently in connection with models of non-Fermi liquids; see, for example, \FaulknerWJ, \EdalatiBI.}   This geometry was obtained
as a solution to D=4+1 Einstein-Maxwell-Chern-Simons theory, and interpolates between
AdS$_5$ at large $r$, and AdS$_3\times R^2$ at small $r$.  The spacetime
is supported by a constant  field strength filling two spatial dimensions.  As discussed in  \DHokerMM,
these solutions include holographic duals to D=3+1 super-Yang-Mills theories, including the maximally 
supersymmetric ${\cal N}=4$ theory,
in the presence of an external magnetic field coupling to their R-current.  Finite
density generalizations of these solutions were obtained and studied, both numerically and analytically, in 
\DHokerBC, \DHokerRZ, \DHokerIJ\ (see also \AstefaneseiDK), and a rich structure was discovered, 
including the existence of a quantum critical point with non-trivial critical exponents.

Given an RG flow, it is interesting to explore how the IR CFT emerges from the structure of correlation
functions computed in the full theory.  One may expect that, in the low energy limit, the correlation 
functions should be describable in terms of an IR CFT, governed by the dual near-horizon geometry.  
It has been pointed out recently in \NickelPR, \HeemskerkHK, and \FaulknerJY\ however, that this 
description of IR behavior may be incomplete when gapless degrees of freedom (or Goldstone
bosons) exist which are supported in the bulk, for finite $r$. These issues can be studied 
quite concretely and explicitly using holographic RG flow solutions.    

In this paper we compute low energy two-point correlation functions for the current and the stress tensor, 
dual to perturbations of  the gauge field and metric of the bulk gravity solution found in  \DHokerMM.  
We work at zero temperature.  The bulk solutions, obtained in \DHokerMM\  involve some numerical 
input.  But as we will see,  enough is known about the asymptotics of the solutions to obtain 
analytic results for the low energy correlation functions. 

On general grounds, we expect that the stress tensor two-point function should be consistent
with the existence of a D=1+1 CFT with a central charge given by the Brown-Henneaux formula \BrownNW.    In particular, we should be able to infer the existence of a pair of Virasoro algebras
generated by the modes of the stress tensor.   In the standard Brown-Henneaux analysis for AdS$_3$,
the Virasoro generators implement coordinate transformations that act nontrivially on the AdS$_3$
boundary.   On the other hand, in our analysis, the stress tensor is defined at the AdS$_5$ boundary.
This leads to a small puzzle, in that the standard asymptotic boundary conditions for AdS$_5$
are only compatible with a finite dimensional group of asymptotic coordinate transformations. So
how can the infinite dimensional algebra of Brown-Henneaux coordinate transformations manifest
itself in terms of the AdS$_5$ stress tensor?   We answer this question by explicit computation.
The key point turns out to be that the Brown-Henneaux coordinate transformations defined in 
the near horizon AdS$_3$ region turn into physical (non pure-gauge) modes when extended
to the AdS$_5$ region.  The relationship between the AdS$_3$ and AdS$_5$ stress tensors
is then found to be precisely what is needed in order for a Virasoro algebra to emerge
at the AdS$_5$ boundary.   Understanding gained from this example should be  useful more generally
in understanding how emergent symmetries  arise in the IR.   

Another set of issues arises when we consider correlation functions of the boundary current $\cJ_\pm$. 
The bulk gauge field is governed by Maxwell and Chern-Simon terms, with
the latter leading to a chiral anomaly for the dual boundary current.   The leading 
long distance behavior of the current correlator is constrained by the need to saturate
the chiral anomaly. To leading order in the IR, the two-point functions $\langle \cJ_- \cJ_\pm \rangle$
are subdominant so that $\langle \cJ_+ \cJ_+\rangle $ by itself must saturate the chiral anomaly.
This forces  $\langle \cJ_+ \cJ_+\rangle $ to have a $1/(x^+)^2$ falloff\foot{Throughout,  $x^\pm$ refer to 
light-cone coordinates along the D=1+1 boundary directions, while $\pm$ refer to the corresponding 
Einstein indices.}  with a specified coefficient, 
the sign of which is mandated by unitarity. (Reversing the sign of the Chern-Simons coupling will 
reverse the chiralities of the currents, but will maintain the sign of this coefficient, in keeping with unitarity.) 
At intermediate momenta, large compared with the AdS$_3$ scale but small compared to the AdS$_5$ 
scale, the situations is reversed: $\langle \cJ_+ \cJ_\pm\rangle $ are now subdominant, and it is
$\langle \cJ_- \cJ_- \rangle$ that saturates the anomaly.  The sign of the coefficient 
in this correlator is now opposite to the one required by ``naive unitarity" in this  intermediate regime.
Unitarity is maintained, however,  in the full magnetic brane solution with AdS$_5$
asymptotics. We show that all these properties are indeed borne out in our holographic calculations.

In the IR behavior of the current correlator $\langle \cJ_+ \cJ_+\rangle $, we also find an infinite series of 
sub-leading terms, whose analogues also appear in the correlators $\langle \cJ_- \cJ_\pm\rangle $.  
We show how these sub-leading terms can be understood
in terms of double trace interactions  generated in the RG flow towards the IR fixed point CFT. Such effects
have been discussed recently in \HeemskerkHK\ and \FaulknerJY.    The double trace interaction involves
a current bilinear, and as usual \WittenUA\ its coupling $\zeta$ is related to the form of mixed 
Neumann-Dirichlet boundary conditions for the gauge field at the AdS$_3$ boundary.  
In fact, as far as the current correlators are concerned,  all the information about how the AdS$_3$ 
geometry is embedded in the magnetic brane solution with  AdS$_5$ asymptotics is contained in the value of the parameter $\zeta$, whose sole dependence is on the Chern-Simons coupling. 

Our computations are conceptually instructive in that they illustrate how to derive new AdS/CFT dictionaries
from old ones.  In particular, the holographic dictionary for a gauge field in AdS$_3$ with both Maxwell
and Chern-Simons terms is subtle (and has not been written down explicitly as far as we know; 
see \KrausNB, \KrausWN\ for the pure Chern-Simons case), due to the presence of several modes with
different growth at the boundary.  On the other hand, the corresponding dictionary for a gauge field
in an asymptotically AdS$_5$ geometry is straightforward.  By taking the low energy limit of the
correlation functions, we can use the latter to derive the rules for the former. 
 
The remainder of this paper is organized as follows.  In section 2 we review the construction of the
boundary stress tensor and current for our theory.  In section 3 we collect needed results on the 
magnetic brane solution interpolating between AdS$_5$ and AdS$_3 \times R^2$.  In section 4 we
compute the current correlators, and discuss their interpretation in section 5.   We  consider the
stress tensor correlators in section 6, and conclude with some comments in section 7. The analytic 
continuation between Minkowski and Euclidean signatures is
relayed to Appendix A, while an evaluation of needed Fourier integrals may be found in Appendix B.

The computations  in this paper can be extended to the case of finite charge density, and the
results will appear separately \toappear.

\newsec{Stress tensor and current definitions}

The action of five-dimensional Einstein-Maxwell-Chern-Simons theory with a negative cosmological 
constant is\foot{The action is for a metric with Minkowski signature. Our metric and curvature conventions 
follow those adopted by Weinberg \weinberg.}
\eqn\aa{ S = -{1\over 16\pi G_5} \int_M \! d^5x \sqrt{g} \left(R+F^{MN}F_{MN}-{12\over L^2}\right) 
+S_{CS} + S_{\rm bndy}}
where the Chern-Simons action is given by
\eqn\ab{S_{CS}={k\over 12\pi G_5} \int_M A \wedge F \wedge F }
In these conventions, the action \aa\ coincides with that of minimal $D=5$ gauged supergravity when
$k = 2/\sqrt{3}$.  In this paper  $k$ will be left unspecified, although we will assume $k \geq 0$; 
there is no loss of generality here, since the sign of $k$ is flipped by the field redefinition $A \rt -A$.     
The contributions denoted by $S_{\rm bndy}$ consist of the usual boundary counterterms needed for a well defined variational principle; their explicit
forms will not be needed here.  We henceforth set $L=1$.  

For the supersymmetric value $k=2/\sqrt{3}$,  this action  is a consistent truncation known to describe all
supersymmetric compactifications of Type IIB or M-theory to AdS$_5$
(see \BuchelGB,  \GauntlettAI, and \GauntlettMA).  This means that solutions of 
\ab\  are guaranteed to be solutions of the full 10 or 11 dimensional field equations 
(although for non-supersymmetric solutions there is no guarantee of stability).  
It also implies that the solutions we find
are holographically dual not just to ${\cal N}=4$ super-Yang-Mills, but to the infinite class of 
supersymmetric field theories dual to these more general supersymmetric AdS$_5$ compactifications. 

The Bianchi identity is $dF=0$, while the field equations are given by,
\eqn\ac{ \eqalign{ 
0 & =   d  \star  F + k F \wedge F
\cr
R_{MN} & =  
4  g_{MN} +{1 \over 3} F^{PQ}F_{PQ} g_{MN} -2 F_{MP}F_N{}^P }} 

We will be considering asymptotically AdS$_5$ solutions, in the sense that the metric and gauge
field admit a Fefferman-Graham expansion \FG.  Introducing a radial coordinate $\rho$, defined
such that the AdS$_5$ boundary is located at $\rho=\infty$,  the metric takes the asymptotic form
\eqn\ad{\eqalign{ ds^2 &= {d\rho^2 \over 4\rho^2} +g_{\mu\nu}(\rho,x) dx^\mu dx^\nu \cr
g_{\mu\nu}(\rho,x) & = \rho \go_{\mu\nu}(x) + \gt_{\mu\nu}(x) + {1 \over \rho}\gf_{\mu\nu}(x)   
+{\ln \rho \over \rho}  \gln_{\mu\nu}(x)  + \cdots  }}
and for the gauge field we have
\eqn\ae{\eqalign{ A & = A_\mu(\rho,x) dx^\mu \cr
A_\mu(\rho,x)& = \Ao_\mu(x)+{1\over \rho} \At_{\mu}(x) +\cdots }}
Here $x^M = (\rho,x^\mu)$, with $\mu = 0,1,2,3$.  

A key point in the construction of the boundary stress tensor \BalasubramanianRE\deHaroXN\ is that the coefficients 
$\gt_{\mu\nu}$ and $\gln_{\mu\nu}$ are fixed by the Einstein equations to be local functionals of the conformal 
boundary metric $\go_{\mu\nu}$. On the other hand, for $\gf_{\mu\nu}$  the  Einstein equations only fix the trace,  
$\tr ({\go}^{-1} \gf)$, to be a local functional of $\go_{\mu\nu}$. 

The boundary stress tensor and current are defined in terms of the variation of the on-shell action
\eqn\af{\eqalign{  \delta S =  \int\! d^4x \sqrt{ \go} \left( {1\over 2}T^{\mu\nu} \delta \go_{\mu\nu} 
+ J^\mu \delta \Ao_\mu \right)}}
In terms of the Fefferman-Graham data the result is  \BalasubramanianRE\deHaroXN\
\eqn\ag{ \eqalign{ 4\pi G_5 T_{\mu\nu}(x) & = \gf_{\mu\nu}(x) +{\rm local}  \cr
2\pi G_5 J_\mu(x) & = \At_\mu(x)+{\rm local}   }}
where indices are  lowered using the conformal boundary metric $\go_{\mu\nu}$. 
The local terms denote tensors constructed locally from $\go_{\mu\nu}$ and $\Ao_\mu$.  In this paper
we are interested in computing two-point correlation functions of operators at non-coincident points. 
For this, we need to compute the stress tensor and current induced at one point by a variation of 
$\go_{\mu\nu}$ and $\Ao_{\mu}$ at a different point.   The local terms in \ag\ do not contribute,
and hence are not needed for computing correlators of operators at distinct points; they instead contribute
to contact terms involving delta functions and derivatives of delta functions.   We henceforth drop the
local terms.  

\newsec{The magnetic background solution}

\subsec{Gravity solution} 

Our background solution is holographically dual to a four-dimensional CFT in a constant 
external magnetic field.  The bulk solution takes the form
\eqn\ah{\eqalign{ ds^2 & = {dr^2 \over L_0(r)^2}+2L_0(r) dx^+ dx^- 
+ e^{2V_0(r)}dx^i dx^i~,\quad\quad i=1,2 \cr
F & = b dx^1 \wedge dx^2 }}
The value of $b$ can be changed by rescaling $x^i$; a convenient choice turns out to be
\eqn\ai{ b= \sqrt{3} }
Inserting the ${\it Ansatz}$ \ah\ into the field equations gives the following the  equations, 
\eqn\aj{\eqalign{& L_0''+2V_0'L_0'+4(V_0''+V_0'^2)L_0=0 
\cr
& 6 L_0^2 V_0'' + 8 L_0^2 (V_0')^2 + 4 L_0L_0'V_0' - (L_0')^2 + 12 e^{-4V_0} =0 
 \cr
& 4L_0^2 (V_0')^2 + 8 L_0 L_0' V_0' -24 + (L_0')^2 + 12e^{-4V_0}=0}}
We note that there is some redundancy in this system of equations, since by using the
derivative of the third equation one can show that one combination of the first two equations is
obeyed identically. 
$L_0$ can be determined in terms of $V_0$ as 
\eqn\ak{ 
L_0(r)^2 =24 e^{-2V_0(r)}\int_0^r \! dr' \int_{0}^{r'}\! dr'' e^{2V_0(r'')}}
The function $V_0$ is determined numerically; see \DHokerIJ.  

\subsec{Asymptotic behavior of the solution}

As $r\rt \infty$ the solution approaches AdS$_5$.  The asymptotics are
\eqn\al{\eqalign{ L_0(r)&= 2(r-r_0)  -{3\ln r \over c_V^2 r} + {\cal O}(r^{-1}) \cr
e^{2V_0(r)} &= c_V(r-r_0) + {\cal O}(r^{-1})  }}
If we scale $x^i$ such that \ai\ is obeyed, and shift $r$ such that $L(0)=0$, then the 
constants $c_V$ and $r_0$ are found numerically to be $c_V \approx 2.797$, $r_0 \approx 0.53$. 
The AdS$_5$ radius is $L_5 = L =1$. 

The coordinate $r$ differs from the coordinate $\rho$ appearing in the Fefferman-Graham expansion.
At large $r$ they are related by
\eqn\ala{ \rho = 4(r-r_0) +{\ell_1 \over r} -{3\over 2 c_V^2 r}(1+2 \ln r) + \cdots}
where  $\ell_1$ is a constant  that can be computed numerically in terms of $L_0$ and
$V_0$  (we will not need its value).  In writing \ala\ we have chosen a convenient rescaling
of the $\rho $ coordinate, so that the 
conformal boundary metric is
\eqn\alb{ \go_{\mu\nu}dx^\mu dx^\nu = dx^+ dx^-  +{c_V \over 4}dx^i dx^i}

As $r \rt 0$ the solution approaches AdS$_3 \times R^2$, with asymptotics 
\eqn\am{\eqalign{ L_0(r) & = 2br +{\cal O}(r^{1+\sigma})\qquad\qquad\qquad 
\sigma = -{1\over 2}+{\sqrt{57}\over 6} \cr
e^{2V_0(r)} &= 1+ 2 r^\sigma +{\cal O}(r^{2\sigma})  }}
The AdS$_3$ radius is 
\eqn\an{L_3 = {1\over b} = {1\over \sqrt{3}} }

Given the AdS$_3$ factor, we can compute the Brown-Henneaux central charge \BrownNW.  
If the transverse $x^{1,2}$ space is infinite, this is really a central charge per unit area.  
If we take the $x^{1,2}$ space to have finite coordinate area $V_2$, then the central charge is
also finite, and given by, 
\eqn\ao{ c = {3L_3 \over 2 G_3} = {3L_3 V_2 \over 2G_5} = {\sqrt{3} V_2 \over 2 G_5 }}

\subsec{CFT interpretation}

The above solution describes an RG flow geometry interpolating between AdS$_5$ at large $r$ and
AdS$_3 \times R^2$ at small $r$.  The CFT interpretation of this flow was given in \DHokerMM.
In the CFT we have  massless charged fermions and bosons propagating in a background magnetic
field.    In the free field limit, the eigenfunctions are Landau levels, localized in the $x^{1,2}$ space, 
but with an arbitrary
momentum $p_3$ parallel to the magnetic field.  Aside from the $p_3$ dependence, the Landau levels 
have a discrete energy spectrum, and at low energies only the lowest Landau level is occupied, leaving
$p_3$ as the only remaining  quantum number.  The low energy theory thus reduces to an  effective 
D=1+1 CFT, corresponding to propagation parallel to the magnetic field lines.  This explains the appearance
of an AdS$_3$ factor in the IR region of the bulk geometry.   As on the gravity side, 
before compactifying the $x^{1,2}$ space we really have a collection of D=1+1 CFTs smeared over
the space, but if we compactify with area $V_2$ then we will obtain a bonafide D=1+1 CFT
with a central charge proportional to $V_2$.    In   \DHokerMM\ this central charge was computed
for free ${\cal N}=4$ Super-Yang-Mills theory and compared with the Brown-Henneaux central charge \ao, 
with the result  $c_{{\cal N}=4} = \sqrt{3\over 4} c_{BH}$.    The agreement up to a numerical factor is similar 
to the agreement up to the factor of $3/4$ in the low temperature entropy of D3-branes \GubserDE. 
As in that case, the solutions under consideration are non-supersymmetric, and there is no 
symmetry protecting the central charge from being renormalized in going from weak to strong coupling.

Another salient point is that  the low energy CFT is populated only by fermionic excitations.
This is because the  lowest Landau level energy for a massless fermion is $E=0$, while for
bosons it is $E \sim \sqrt{B}$.  The existence of the fermion zero mode is due to the negative
Zeeman energy associated with the fermion spin aligning with the magnetic field, which precisely cancels
the kinetic energy.   Alternatively, this is a consequence of the index theorem for the Dirac operator. 
As a consequence, at energies small compared to $\sqrt{B}$, only the lowest fermionic Landau
level participates, and so the CFT has only fermionic excitations. Of course, since this is a 
D=1+1 CFT there is not really a sharp distinction between fermions and bosons.

In the remainder of this paper we use the AdS/CFT correspondence to compute  
CFT correlation functions  at strong coupling.

\newsec{Current correlators}

To compute the current two-point function we proceed by solving the Maxwell equations
with a specified boundary condition $\Ao_\mu(x)$.    Given the solution, we read off the induced
current\foot{Throughout, we will denote the current
operator by $\cJ^\mu$ and its expectation value by $J^\mu$; and similarly for the stress tensor
operator $\cT^{\mu \nu}$ and expectation value $T^{\mu \nu}$.} $J^\mu(x)$ from \ag, and then use this result to extract the correlator according 
to the formula for first order perturbation theory in $\Ao_\nu$,
\eqn\ba{ J^\mu(x) = i   \int\! d^4 y  \sqrt{\go} \langle \cJ^\mu(x) \cJ^\nu(y) \rangle \Ao_\nu(y) }
This formula derives from the Minkowski signature functional integral, whence the 
presence of the prefactor of $i$; the analytic continuation to the Euclidean version 
of this formula will be derived in Appendix A.

The effective CFT that we wish to probe inhabits the $x^\pm$ directions.  For this reason, 
we will only consider correlators of $\cJ^\pm$.   Similarly, working in momentum space, we
will only consider nonzero momenta $p_\pm$, so that in position space we are integrating the
current operators over $x^{1,2}$. 

The two-point function can be computed by working to first order in  $\Ao_{\mu}$, and so 
we only need to solve the field equations to linear order around the background solution.   
In general, the gauge field perturbations can mix at linear order with metric perturbations.
However, if the linearized gauge fluctuations have polarization restricted to $A_\pm$, and
have no dependence on $x^{1,2}$, then it is easy to check that there is no mixing.  Hence we can
consistently set the metric perturbation to zero in this computation, and we just 
need to solve the linearized Maxwell-Chern-Simons equation.  

\subsec{Plane wave expansion}

We choose the gauge $A_r=0$, where the background gauge field obeys,
\eqn\bc{ dA_0 = b dx^1 \wedge dx^2 }
We consider a plane wave perturbation with fixed momenta $p_\pm$ in the $x^\pm$ directions, 
so that the full gauge field takes the form,
\eqn\bb{\eqalign{  A &= A_0 + a_+(r,p_\pm) e^{ipx} dx^+ + a_-(r,p_\pm) e^{ipx}  dx^-
}} 
Throughout, we will use the following notations,
\eqn\bba{\eqalign{ px & = p_+ x^+ +  p_- x^-\cr p^2 & = p_+ p_- }}
Substituting \bb\ into $d\star F + k F\wedge F=0$, we keep only terms linear in $a_\pm$.  
The reduced equations take their simplest form 
if we define the combinations,
\eqn\bez{\eqalign{ \ep_p  &= p_- a_+ + p_+ a_- \cr   \ep_m &= p_- a_+ - p_+ a_-}}
Since the background metric is unperturbed in this computation, we will
drop the $0$ subscript on $L_0$ and $V_0$. The reduced field equations are,
\eqn\bg{\eqalign{& (Le^{2V} \ep_m ')' - {4k^2 b^2 \over Le^{2V}} \ep_m -{2e^{2V} \over L^2} p^2 \ep_m =0 \cr
& Le^{2V} \ep_p ' -2kb \ep_m =0 }}

For general $p_\pm$ these equations must be solved numerically.  However, we can make analytical
progress by focussing on low energy correlators corresponding to the Euclidean  region,
\eqn\bh{ 0 < p^2 \ll 1}
This is the regime in which we expect to probe the  IR  D=1+1 CFT.   

To proceed, we employ a standard approach in this context, that of a matched asymptotic expansion
(see for example \DHokerIJ).
We consider two overlapping regions that together cover the space, a near region and a far region, 
in each of which we can solve the equations analytically.
Under the assumption \bh\ these regions overlap  in a parametrically large region, and by matching
the asymptotics there, we obtain a  solution valid throughout the entire space. 

It is useful to note that the functions $V$ and $L$ appearing in the equations \bg\ depend 
on no free parameters (assuming that we have set $b=\sqrt{3}$), and so the transition between 
their small $r$ near horizon behavior and large $r$
asymptotic behavior occurs for $r \approx 1$. 

\subsec{Near region}

The near region is defined as $r \ll 1$.    In this region we can use the leading small $r$ asymptotics
in $V$ and $L$,
\eqn\bi{ e^{2V}=1 \qquad\qquad\qquad L = 2br }
so that the equations \bg\ become 
\eqn\bj{\eqalign{&  r^2 \ep_m '' +  r \ep_m'  -  k^2 \ep_m -{p^2 \over 12 b r }  \ep_m =0 \cr
& r \ep_p ' -k \ep_m =0 }}
These equations are equivalent to the three-dimensional Maxwell-Chern-Simons equations
$d\star F + 2kb F =0$, with the metric given by AdS$_3$.   Since we are assuming that $p^2 >0$, 
the solution for $\ep_m$ that is smooth at $r=0$ is given by a modified Bessel function,
\eqn\bk{ \ep_m = {2 \sin (2\pi k) \over \pi} C K_{2k}\left( \sqrt{p^2 \over b^3 r}\right)}
The integration constant $C$  has been chosen so as to make the large 
$r/p^2 \to \infty$ asymptotics simple, namely
\eqn\bm{ \ep_m \sim C \left[{ \left( {p^2\over 4b^3}\right)^{-k} \over \Gamma(1-2k)}r^k 
-  {\left({p^2\over 4b^3 }\right)^{k} \over \Gamma(1+2k)}r^{-k} \right] }
Note that the region of large $r/p^2$ overlaps with the near region $r\ll 1$ 
because we are assuming $p^2 \ll 1 $.  The $r/p^2 \to \infty$ asymptotics of $\ep_p$ follow from \bj,
\eqn\bn{  \ep_p \sim C \left[{ \left( {p^2\over 4b^3}\right)^{-k} \over \Gamma(1-2k)}r^k 
+  {\left({p^2\over 4b^3}\right)^{k} \over \Gamma(1+2k)}r^{-k} \right]  + 2p_+ p_- \lambda }
where $\lambda$ is an independent integration constant, chosen in a convenient manner. 
The $r/p^2 \to \infty$ asymptotics of the original field $a_\pm$ is then readily found, 
\eqn\bo{\eqalign{ a_+ & \sim  { C \left(  {p^2\over 4b^3}\right)^{-k} \over \Gamma(1-2k)\, p_-} 
 r^k+ p_+ \lambda  \cr
 a_- & \sim  { C \left( {p^2\over 4b^3}\right)^{k} \over \Gamma(1+2k)\, p_+}  r^{-k} + p_- \lambda }}

\subsec{Far region}

The far region is defined such that we can neglect the momentum dependent term in \bg.   
This is  valid provided $p^2/r \ll 1$.  The equations \bg\ then reduce to the following equations
for $a_\pm$
\eqn\bp{\eqalign{ Le^{2V} a_+' -2kb a_+ & =-2kb  p_+   \lamt \cr 
 Le^{2V} a_-' +2kb a_- & = 2kb p_-  \lamt}}
where $ \lamt$ is an integration constant.   We write the solutions as,
\eqn\bq{\eqalign{ a_+ &= (\azero_+  - p_+  \lamt )e^{2kb\psi(r)} + p_+  \lamt  \cr
a_- &= (\azero_-  - p_-  \lamt )e^{-2kb\psi(r)} + p_-  \lamt }} 
where the function $\psi (r)$ is familiar from \DHokerIJ, and is defined by,
\eqn\br{ \psi(r)= \int_\infty^r {dr' \over L(r')e^{2V(r')}}}
and $\azero_\pm$ are new integration constants. 
The asymptotics of $\psi$ may be evaluated in terms of those of $L$ and $V$, and we find
\eqn\bs{\eqalign{r \rt 0\quad\quad\quad &  \psi(r) = {\ln r \over 2b} +\psi_0+ {\cal O}(r^\sigma) \cr
r \rt \infty \quad\quad\quad  &  \psi(r) = -{1\over 2 c_V r}  + {\cal O}(r^{-2} )}}
where $\psi_0$ is a constant that may be determined numerically to be $\psi_0 \approx 0.2625$.
The asymptotics of $a_\pm$ are then found to be
\eqn\bt{\eqalign{ r \rt 0\qquad\quad\quad &  a_+(r) =  (\azero_+  - p_+  \lamt )e^{2kb\psi_0} r^k + p_+  \lamt \cr
        &  a_-(r) =  (\azero_-  - p_-  \lamt)e^{-2kb \psi_0} r^{-k} + p_- \lamt  \cr
 r \rt \infty \qquad\quad\quad &  a_+(r) =\azero_++{1\over 4r} \atwo_+  \cr
    &  a_-(r) =\azero_- +{1\over 4r} \atwo_-  }}
where,
\eqn\bu{\eqalign{ \atwo_+ & =-{4kb \over c_V}  (\azero_+  - p_+  \lamt ) \cr
  \atwo_- & = {4kb \over c_V}  (\azero_-  - p_-  \lamt ) }}
Using the asymptotic relation $\rho=4r$ of \ala, we see that the asymptotic expansion of \bt\
agrees with the expansion appearing in \ae, so we can  use \ag\ to read off (the Fourier 
transform of) the current as,
\eqn\bva{ J_\mu = {1\over 2\pi G_5} \atwo_\mu }

\subsec{Matching}
 
The matching region is defined by $ p^2 \ll r \ll 1$, which overlaps both the near and far regions. 
In this region we demand agreement between the $r/p^2 \rt \infty$ asymptotics of the near region solution, 
and the $r \ll 1$ asymptotics of the far region solution.   Equating the expressions in \bo\
with those in the top two lines of \bt\ we find that $\lamt =\lambda$, along with two equations
determining $C$ and $\lambda$ in terms of $\azero_\pm$.   A bit of algebra then leads to the following
expressions for $\atwo_\pm$:
\eqn\bv{\eqalign{\atwo_+ &= -{4kb \over c_V} {\azero_+ \over \left[ 1- \zeta \, p^{4k}\right]} 
+ {4kb \over c_V}{p_+ \over p_-}  {\azero_- \over \left[ 1- \zeta \, p^{4k}\right]}\cr
\atwo_- &= +{4kb \over c_V} {\azero_- \over \left[ 1- \zeta^{-1} \, p^{-4k}\right]} 
- {4kb \over c_V}{p_- \over p_+}  {\azero_+ \over \left[ 1- \zeta^{-1} \, p^{-4k}\right]} }}
where
\eqn\bw{ \zeta = \zeta (k) = {\Gamma(1-2k) \over \Gamma(1+2k)}{e^{4kb \psi_0} \over (4b^3)^{2k}}  }
is a function of $k$, and of the (fixed) characteristics of the background solution. Note that,
extending \bw\ to negative values of its argument,  we have $\zeta ^{-1} = \zeta (-k)$.

\subsec{Correlation functions: IR behavior}

Since we are restricting to zero momentum along $x^{1,2}$, it is convenient to define a two-dimensional
current, $\cJh_\pm$, by integrating over $x^{1,2}$,
\eqn\bx{ \cJh_\pm = {1 \over 4} c_V  V_2 \cJ_\pm }
Using \bva\ it can be expressed as
\eqn\by{ \cJh_\pm = {c_V c\over 4\pi b} \atwo_\pm }
where $c$ is the Brown-Henneaux central charge \ao.

Using the analytic continuation formulas derived in Appendix A, we  read off the correlators from \bv,
expressed in terms of Euclidean momenta,
\eqn\bz{\eqalign{ \langle \cJh_+(p) \cJh_+(-p) \rangle 
 &  = {kc \over 2\pi}  {p_+ \over p_-} + {kc \over 2\pi} {p_+ \over p_-} 
\left ( { \zeta \, p^{4k} \over  1- \zeta \, p^{4k}} \right ) \cr
 \langle \cJh_-(p) \cJh_-(-p) \rangle & = {kc\over 2\pi}  {p_- \over p_+} 
\left ( { \zeta \, p^{4k} \over  1- \zeta \, p^{4k}}  \right ) \cr
 \langle \cJh_+(p) \cJh_-(-p) \rangle
 &  =- {kc\over 2\pi} \left ( { \zeta \, p^{4k} \over 1- \zeta \, p^{4k} } \right )
}}
In writing $ \langle \cJh_+ \cJh_- \rangle$ we used the freedom to add a contact term, and for the
same reason, the correlator can be extracted from either the top or bottom line of \bv, with the
same result modulo contact terms. 

The leading non-analytic long-distance behavior, corresponding to small momenta $\zeta p^{4k} \ll 1$ 
is given by, 
\eqn\ca{  \langle \cJh_+(p) \cJh_+(-p) \rangle = {kc \over 2\pi} {p_+ \over p_-}+\cdots }
Fourier transforming this to position space gives (see Appendix B for formulas),
\eqn\cb{  \langle \cJh_+(x^+) \cJh_+(y^+ ) \rangle = -{kc \over 2\pi^2}{1\over (x^+-y^+)^2}+\cdots}
With the conventions adopted here, the minus sign here is actually the one 
required by unitarity, as confirmed in Appendix C.
Keeping the full momentum dependence in \bz\ instead leads to an expansion,
\eqn\cc{  \langle \cJh_+(x) \cJh_+(0 ) \rangle 
= -{kc \over 2\pi^2}{1\over (x^+)^2}\left(1 + \sum_{n=1}^\infty { a_n \over (x^+x^-)^{2nk}} \right)   }
where we have used translation invariance to set $y=0$.
Similarly,  the remaining correlation functions have the structure,
\eqn\cd{\eqalign{   \langle \cJh_-(x) \cJh_-(0 ) \rangle & =  {kc \over 2\pi^2}{1\over (x^-)^2} 
\sum_{n=1}^\infty { b_n \over (x^+x^-)^{2nk}}  \cr 
   \langle \cJh_+(x) \cJh_-(0) \rangle & = {kc \over 2\pi^2}\sum_{n=1}^\infty { c_n \over (x^+x^-)^{2nk}}  }}

In our derivation of these results we proceeded under the condition that $0 < p^2 \ll 1$. 
On the one hand, we have neglected corrections of the type $p^2/r$ in the region $r >1$.
On the other hand, 
in \bz\ we wrote the full functional dependence on $p^2$.   We now make a comment
regarding in what sense this is meaningful.  We expect that the full answer, not assuming $p^2 \ll 1$,
would add to the expressions in \bz\ terms suppressed by $p^2$ compared to the leading terms. 
Since the sub-leading terms in \bz\ are down by powers of $p^{4nk}$, it is clear that for any
fixed $k$ we should only keep a finite number of terms in the expansion, namely those for which
$2nk <1$.      By the same token, by taking $k$ sufficiently small we can always arrange for an
arbitrarily large number of terms in the expansion to dominate the terms we have been neglecting, and
so in this sense the full expressions in \bz\ are meaningful.

\subsec{Correlation functions: UV behavior}

The UV behavior of the correlators is obtained by solving equations \bg\ for $p_+ p_- \gg 1$.
Although the short-distance behavior is not central to the main theme of this paper, we 
shall include it briefly here for the sake of completeness, and verification of overall signs.
The UV limit of the correlators reduces to those of pure AdS$_5$. We can therefore
set $k=0$,  $e^{2V}=c_Vr$, and $L=2r$ in \bg. The solution for $\ep_m$ which is smooth as $r \to 0$ 
is proportional to the Bessel function $K_1 (2p/\sqrt{r})$.  Extracting its large $r$ behavior, we 
obtain the following expression for the (non-local part of the) current, 
\eqn\ze{ 4 \pi G_5 J_\pm = - p_+^2 (\ln p^2) \azero_\mp  }
The UV limits of the current correlators are then found as follows (see Appendices A
and B for their evaluation),
\eqn\zg{ \eqalign{ \langle \cJ_\pm (p) \cJ_\pm  (-p)\rangle & =    - { p_\pm^2 \ln p^2 \over 4 \pi G_5}  \cr
\langle \cJ_\pm(x) \cJ_\pm (0)\rangle  & =    - {1\over 2 \pi^2 G_5 } { 1\over (x^\pm)^2 (x^+ x^-)}  }}
Although $\cJ_\pm$ are dimension 3 operators in the UV, the correlators  in \zg\ have a $1/x^4$ falloff
because they have been integrated over $x^{1,2}$. 
Comparing the signs of the position space correlators in the IR of \cb, and in the UV of
\zg, we find agreement between the signs of the leading coefficients, as is required 
by unitarity of the low energy sector, and of the full UV theory.
Thus, there is no sign change generated by the RG flow.

\newsec{Discussion of current correlators} 

In this section, we discuss various physical aspects of our results for the current correlation functions.   

\subsec{Chiral anomaly and unitarity}

The chiral anomaly equation for the current $\cJh_\pm$ may be deduced directly from \bv\ in momentum space,
and is given by,
\eqn\bvan{
p_+ \cJh _- + p_- \cJh_+ = {k c \over \pi} \left ( p_+ a_- ^{(0)} - p_- a_+ ^{(0)} \right )}
The anomaly is free of higher order $p^{4k}$ momentum corrections, as expected.
The leading IR behavior of the current correlators \cc\ and \cd\ is constrained 
by the chiral anomaly \bvan.  In the bulk, this anomaly equation in the IR
arises due to the fact that the D=4+1  Chern-Simons terms $k \int \!A\wedge F \wedge F$ reduces
to a D=2+1 dimensional Chern-Simons term $kb  \int \! A \wedge F$. Bulk Chern-Simons terms
are directly related to anomalies of boundary currents.  This is manifest here through the fact
that the anomaly in \bvan\ is linear in $k$, with parity reversing the sign of $k$ and the roles of 
the $\pm$ chiralities. It is also manifest in the structure of \bv\ in which chiralities 
are reversed under $ k \to -k$ upon using the fact that $\zeta (-k) = \zeta (k)^{-1}$.
Therefore, without loss of generality, we continue to make the choice $k>0$,
the case of $k<0$ being obtained by reversing chiralities.

We now use our explicit results of \bv\ and of the correlators \bz\ to analyze the interplay
between the chiral anomaly and unitarity of the IR effective CFT. To leading order in the IR, 
the two-point functions $\langle \cJh_- \cJh_- \rangle$ and $\langle \cJh_+ \cJh_- \rangle$
are subdominant, so that $\langle \cJh_+ \cJh_+\rangle $ must saturate the chiral anomaly by itself.
This forces  $\langle \cJh_+ \cJh_+\rangle $ in \cb\ to have a $1/(x^+)^2$ falloff with a specified coefficient, 
$-kc/(2 \pi ^2)<0$, the sign of which is mandated by unitarity.

Now consider the regime  $\zeta p^{4k} \gg 1 $ in \bz. To the extent that \bz\ represent the
correlators of some D=1+1  QFT, this limit probes the short distance regime of this QFT.  
 One sees that  the situation as regards the chiral anomaly is  now reversed: the correlators $\langle \cJh_+ \cJh_+\rangle $
and $\langle \cJh_+ \cJh_- \rangle$ are now subdominant, and it is the correlator 
$\langle \cJh_- \cJh_- \rangle$ which saturates the  anomaly.  The coefficient of
$1/(x^-)^2$ in this correlator is now given by $kc/(2\pi^2)$, and is opposite to the one required 
by ``naive unitarity" in this  regime.  This is telling us that the expressions in  \bz\ cannot by themselves be interpreted
as the correlators of some unitary QFT; there must be corrections that set in in the UV to maintain
unitarity.   In the present context we know precisely what these corrections represent.  We
have already noted that the approximations leading to \bz\ break down for $p \sim 1$.
For momenta larger than this, the correlators will start to ``see" the AdS$_5$ region.   Unitarity in the full magnetic brane solution with 
AdS$_5$ asymptotics is maintained, as may be seen explicitly, for example, 
from the UV limit of the correlators given in \zg. These facts fit together nicely.

\subsec{Sub-leading terms and double trace operators}

Next, let us consider the interpretation of the sub-leading terms.   In the near horizon geometry
that governs the IR physics, the bulk Maxwell field obeys a D=2+1 dimensional Maxwell-Chern-Simons
equation.  As is well known, the Chern-Simons term gives the gauge field a mass proportional to $k$. 
On the boundary, this corresponds to an operator with a $k$-dependent scaling dimension, which accounts
for the $k$ dependent powers appearing in the correlation functions.   On the other hand, the fact
that we have an infinite series of different powers appearing in the correlators indicates that these
operators do not have a definite scaling dimension, and that the theory is not scale invariant.  
The reason for this can be explained in terms of modified boundary conditions and double trace operators.

To understand this let us examine the relation between the behavior of the gauge field near the
UV AdS$_5$ boundary versus at the near horizon IR AdS$_3$ boundary.    In particular, we focus
on the relation between the ``source" and ``vev" terms in the two regions.    At the AdS$_5$ boundary
we are using the standard AdS/CFT dictionary, which identifies the source as $\azero_\pm$ and
the vev as $\atwo_\pm$.  Now, at the boundary of the AdS$_3$ region, which corresponds to the
matching region $|p_+ p_-| \ll r \ll 1$,  a generic solution of the field equations has the
expansion
\eqn\ce{ \eqalign{ a_+ & \sim  C_+ r^k + p_+ \lambda \cr 
 a_-  & \sim  C_- r^{-k}  + p_- \lambda }}
In terms of these quantities, it is not immediately obvious how to relate source and vev terms
to $C_+$, $C_-$ and $\lambda$.  However, since the field equations in the far region relate
these coefficients to the data at the AdS$_5$ boundary, we can use the AdS$_5$/CFT$_4$ 
dictionary to answer this question.    

First consider the current.   From \bt-\bv, we see that
there is a simple relation $\cJ_\pm \propto C_\pm$, between the current and the coefficients
of the $r^{\pm k}$ terms in the matching region.  Thus the current can be immediately read off
from the AdS$_3$ near-boundary behavior shown in \ce.   

For the source, comparing \bt\ to \ce\ now leads to the relations
\eqn\cf{\eqalign{ \azero_+ & = e^{-2kb \psi_0} C_+ +p_+ \lambda \cr
 \azero_- & = e^{2kb \psi_0} C_- +p_- \lambda }}
We can think of this as a version of mixed Neumann-Dirichlet boundary conditions at the AdS$_3$
boundary.  Note that all the information about the far region is contained in the factor
$e^{2kb\psi_0}$; if we imagine a more general family of geometries interpolating between
the AdS$_3$ and AdS$_5$ regions we can think of  $e^{2kb\psi_0}$ as being a variable parameter that controls
the form of the IR boundary condition.    

In AdS/CFT it is well known that considering mixed boundary conditions corresponds to adding
double trace interactions to the Lagrangian of the boundary CFT \WittenUA.   These double trace interactions
induce a nontrivial RG flow in the field theory, and so correlation functions acquire a nontrivial 
dependence on momenta, exhibiting the interpolation between the UV and IR fixed points.  The 
explicit form of such correlators is determined using large $N$ factorization, and takes
the form shown in \bz.   

The fact that we generate double trace interactions in the approach to the IR fixed point is not
surprising.  We start from the UV CFT at the AdS$_5$ boundary, and then add an external magnetic
field that introduces a scale.    The theory then undergoes an RG flow to the IR, generating in the
process all possible operators allowed by symmetry and large $N$ counting.   The appearance
of double trace interactions in holographic RG flows has been discussed in detail in recent 
papers  \HeemskerkHK\FaulknerJY, and a discussion of boundary conditions for gauge fields in
AdS is found in  \MarolfND.

We now consider in more detail the form of the double trace interactions.  To do so, we first
define the CFT in the absence of double trace terms.   We write the
current operators in the IR CFT as
\eqn\cfa{ \eqalign {  \cJh_+ &=  \sqrt{kc \over 2\pi} \, \p_+ \phi + \sqrt{kc \zeta \over 2\pi} \p_+ \cO  \cr 
\cJh_- & =  \sqrt{kc  \zeta\over 2\pi}\p_- \cO  }}
Here $\phi$ is a free boson, whose momentum space  two-point function is, 
\eqn\cfb{ \langle  \phi(p)    \phi(-p)  \rangle  ={1 \over p^2} }
while $\cO$ is a scalar operator of dimension $(k,k)$, with two-point function,
\eqn\cfc{\langle \cO(p)\cO(-p)\rangle = (p^2)^{2k-1}  }
The mixed correlator $\langle \phi \cO \rangle$ is assumed to vanish.
Given these two-point functions, if we now add to the CFT Lagrangian the 
double trace term $ \zeta  \p_+ \cO \p_- \cO $, and use large $N$ factorization, it is easy to 
see that we recover the correlators displayed in \bz.  On the one hand, since $\p_+ \cO \p_- \cO $ is an operator 
of total scaling dimension $2k+2$, it is irrelevant in the RG sense for any $k>0$.  
On the other hand, the total scaling dimension $2k$ of the operator $\cO$ itself is below the 
Breitenlohner-Freedman bound (whose value is 1 for the asymptotic AdS$_3$ near-horizon region) 
when $0<k<1/2$, thus suggesting the existence of an instability in this range of $k$.
Remarkably, the same range of $k$ is singled out in the presence of non-zero charge 
density in \toappear.

To summarize, the content of the IR CFT in the sector dual to the bulk gauge field consists of a 
boson $\phi$ together with the operator $\cO$ of dimension $(k,k)$.  
The theory contains a double trace interaction  $ \zeta  \p_+ \cO \p_- \cO $.   From the point of view of
the IR theory, $\zeta$ can be viewed as free parameter.  It takes a definite value upon embedding
the theory in a specific UV CFT, as in \bw.

\subsec{The $k \to 0$ limit of the sub-leading terms}

In the limit $k \to 0$, the contribution of the Chern-Simons term in the action vanishes, and 
the chiral anomaly is cancelled. But the structure of the current correlators is non-trivial.
In fact, in the small $k$ limit, it becomes reliable to keep all the sub-leading expansion
terms in the current correlators \bz\ since further corrections in integer powers of $p^2$
will now be small compared to all expansion terms, as long as $p^2 \ll 1$. The $k \to 0$ limits 
of the current correlators are as follows, 
\eqn\bzzz{\eqalign{ 
\langle \cJh_+(p) \cJh_+(-p) \rangle 
 &  =  - {c \, p_+ \over 2\pi p_-} \, { 1 \over \zeta '(0) + 2 \ln (p^2)}  \cr
 \langle \cJh_-(p) \cJh_-(-p) \rangle & = - { c \, p_- \over 2\pi p_+ }  \, { 1 \over \zeta '(0) + 2 \ln (p^2)}  \cr
 \langle \cJh_+(p) \cJh_-(-p) \rangle
 &  = {c\over 2\pi} \, { 1 \over \zeta '(0) + 2 \ln (p^2)}
}}
where $\zeta ' (0$ is the derivative in $k$ at $k=0$ of the function $\zeta (k)$ of \bw.
To leading order in small $p^2$, the $\zeta '(0)$ term may in fact be omitted. A pole
appears at a finite value of $p^2$ in \bzzz, but this effect is of course beyond the 
range of validity of \bzzz.

\subsec{Relation to Luttinger liquid theory}

Before turning to stress tensor correlators, let us mention also the connection between our
results and those appearing in the Luttinger liquid approach to interacting condensed matter
systems in D=1+1 (see, e.g., \giamarchi.)   Such systems are studied using bosonization methods.   Linearizing
around the Fermi surface, one obtains fermions with a relativistic type dispersion relation,
and these can be bosonized in a standard fashion.   In the simplest setup corresponding
to spinless fermions interacting via four-fermi terms, the charge density operator is bosonized as
\eqn\cg{ \rho  \sim \p_x \phi + \Big[ e^{2ik_F x + 2i \phi} +{\rm H.C.}\Big]}
where $\phi$ is a free compact boson, whose radius depends on the four-Fermi couplings.  
We can compare this with the density operator $\rho  = \cJh_+ + \cJh_-$ obtained from \cfa. 
In both cases, the addition of the extra operators, beyond the $\phi$ derivative piece, leads
to sub-leading terms in the density-density correlators.  One obvious difference between \cfa\ and \cg\ is the
presence of the $e^{2i k_F x}$ factor in \cg, which corresponds to a process where a
fermion is taken from one side of the Fermi surface to the other.  There is no analog
of such a factor in our case because we are working at zero density.    Correlators at finite
density, and their connections with Luttinger liquids, will be presented in \toappear. 
See \HungQK\MaityZZ\BalasubramanianSC\ for other discussions of holography and Luttinger liquids.
 
\newsec{Stress tensor correlators}

The strategy for computing stress tensor two-point functions is very similar to that employed in the last
section for the current correlators.  We  solve the linearized Einstein equations with specified
boundary condition $\go_{\mu\nu}$.  Given this solution we compute $T^{\mu\nu}(x)$, and
then use this to read off the two-point function via,
\eqn\da{ T^{\mu\nu}(x) ={i\over 2} \int\! d^4y \sqrt{\go} 
\langle \cT^{\mu\nu}(x) \cT^{\alpha\beta}(y)\rangle \go_{\alpha\beta}(y)}
 To isolate the effective D=1+1 CFT, we restrict the graviton polarizations to the $x^\pm$
components, and restrict the momenta to $p_\pm$.  In the regime $0< p_+ p_- \ll 1$ we
can solve the problem analytically by using a matched asymptotic expansion.

More precisely, we consider the field configuration
\eqn\db{\eqalign{ ds^2 &= {dr^2 \over L ^2}  + 2L dx^+ dx^- +M (dx^+)^2 + N(dx^-)^2 + e^{2V_0}dx^i dx^i \cr
F & = b dx^1 \wedge dx^2 }}
with
\eqn\dba{\eqalign{ L(r) & = L_0(r) + L_1(r) e^{ipx} \cr
 M & =  M_1(r) e^{ipx} \cr
N & =  N_1(r) e^{ipx} }}
The perturbations are therefore given by $L_1$, $M_1$, and $N_1$, and we proceed by
solving the Einstein equations to linear order in these functions.   In writing \db\ we have used 
the fact that with this choice of metric perturbation there is no
mixing with gauge field fluctuations, and so the latter can be set to zero.  We have also made a
choice of coordinates which is convenient for the analysis of the far region equations. 

The perturbation $L_1$ will be set to zero when we come to the far region, although it is 
helpful to retain it for the time being.   It is only necessary to consider the $L_1$ perturbation
if we are interested in computing a two-point function involving $T_{+-}$.   However, all such
correlators are pure contact terms, vanishing when the two operators are at distinct points.  
This is a reflection of the trace anomaly of the IR  D=1+1  CFT, stating that $T_{+-}$ can
be expressed locally in terms of the conformal boundary metric; namely, it is just proportional
to the Ricci scalar.  For this reason, up to contact terms, all low energy correlators can be
accessed while setting $L_1=0$.

\subsec{Near region}

As in the last section, the near region is defined by $r \ll 1$, where we can set $L_0=2br$ and
$e^{2V_0}=1$.  It is instructive to solve the linearized Einstein equations in a two-step process. 
We first solve the equations in a different coordinate system than in \db, and then perform
the appropriate coordinate transformation to put the solution in the form of \db. 
In particular, we first  consider the perturbed field configuration
\eqn\dbb{\eqalign{ ds^2 &= ds_B^2 +\Big[  h_{++}(r) (dx^+)^2+  2h_{+-}(r)dx^+ dx^- 
+h_{--}(r) (dx^-)^2 \Big] e^{ipx} \cr
F & = b dx^1 \wedge dx^2 }}
where $ds_B^2$ now denotes the background metric \ah. The general solution of the 
linearized Einstein equations, given the {\it Ansatz} \db, is 
\eqn\dc{\eqalign{ h_{++}(r)&=s_{++}r+t_{++} \cr 
 h_{--}(r)&=s_{--}r+t_{--} \cr 
 h_{+-}(r)&=s_{+-}r+t_{+-} }}
where $s$ and $t$ are independent of $r$ with,
\eqn\dd{\eqalign{
t_{+-}&= {1 \over 24 b}\left(p_+^2 s_{--}+p_-^2 s_{++}-2 p_+ p_-  s_{+-}\right)\cr
p_-  t_{++}&= p_+ t_{+-} \cr
p_+ t_{--}&=p_-  t_{+-}  
}}
The structure of the solution is easy to understand.  The equations governing $h_{\mu\nu}$ 
are just those of D=2+1 gravity expanded around AdS$_3$, and solutions are thus locally
pure gauge.   Furthermore, the coefficients $s_{\mu\nu}$ and $t_{\mu\nu}$ can be identified
in terms of the AdS$_3$/CFT$_2$ correspondence:    $s_{\mu\nu}$ represents the perturbation
of the conformal boundary metric, and $t_{\mu\nu}$ is proportional to  the boundary stress tensor.   
The top line
of \dd\ is then identified as the linearized trace anomaly for the stress tensor (the right hand side
is proportional to the linearized Ricci scalar), while the bottom two lines are the equations
representing conservation of the stress tensor.     

We  now change coordinates to put the perturbed solution of \dc\ and \dd\ in the form of \db.  In fact,
for what follows we only need the resulting solution in the region $ p^2 \ll r \ll 1$, for which  
we find
\eqn\de{\eqalign{ L_1(r) &  \sim t_{+-} \cr
M_1(r) & \sim  s_{++}r + t_{++} \cr
N_1(r) & \sim s_{--}r + t_{--}  }}
with
\eqn\df{\eqalign{ t_{++} & = {1\over 24b} \left( p_+ p_- s_{++} + {p_+^3 \over p_-} s_{--}\right)\cr
 t_{--} & = {1\over 24b} \left( p_+ p_- s_{--} + {p_-^3 \over p_+} s_{++}\right) }}
The relations \df\ follow from \dd\ with $s_{+-}=0$.   All dependence on $s_{+-}$ is absorbed by the 
coordinate transformation.    Note that $t_{+-}$ in \de\ is arbitrary; this is
a reflection of the freedom to shift the radial coordinate $r$ while preserving the gauge choice \db. 

\subsec{Far region}

To obtain the far region equations we assume $r \gg p^2$, so that momentum dependent
terms in the field equations can be dropped.  We will now take 
\eqn\dg{ L_1 =0}
This is required by the field equations once we demand that there be no perturbation of the field $V$.
As explained above, setting $L_1 =0$ is permitted when computing the non-contact part of the
correlators. 

 Substituting \db\ into the linearized
Einstein equation, and dropping the momentum dependent terms,  we find the following equations
\eqn\dh{\eqalign{ M_1''+2V_0' M_1 '+4(V_0''+V_0'^2)M_1 & =0  \cr
 N_1''+2V_0' N _1 '+4(V_0''+V_0'^2)N_1 & =0 }}
We now note that $M_1$ and $N_1$ obey the  same equation as obeyed by $L_0$ in the top line
of \aj.   Thus, one solution is given by $L_0$ itself, which is easily seen to correspond to 
acting on the background solution with a  coordinate transformation of the form $x^{+} \rt x^+ + \epsilon x^-$, 
or the equivalent with $x^+ \leftrightarrow x^-$.  The second linearly independent solution
was obtained in  \DHokerIJ\  (see section 5.4 of that paper) and denoted by $L_0^c$.  It is given explicitly by,
\eqn\di{ L_0^c(r) = L_0(r)  \int_\infty^r {dr' \over L_0(r')^2 e^{2V_0(r')}}}
This function has the following asymptotics,
\eqn\dj{\eqalign{ r \rt 0  \quad\quad\quad\quad & L_0^c \sim -{1\over 2b} \cr
 r \rt \infty   \quad\quad\quad\quad & L_0^c \sim  -{1\over 4 c_V r} }}
The solutions for $M_1$ and $N_1$ are arbitrary linear combinations of $L_0$ and
$L_0^c$,
\eqn\dk{\eqalign{ M_1(r)  & = 2 L_0(r) \go_{++}  -c_V L_0^c(r) \tilde{g}^{(4)}_{++}  \cr
 N_1(r)  & =2L_0(r) \go_{--} -c_V L_0^c(r) \tilde{g}^{(4)}_{--}  }}
We have labelled  the coefficients in a convenient manner, noting that the $r \to \infty $  asymptotics are
given by,
\eqn\dl{\eqalign{M_1(r) & \sim \left( 4r + \cdots\right) \go_{++}   
+ {1\over 4r} \tilde{g}^{(4)}_{++}  \cr 
 N_1(r) &\sim \left( 4r + \cdots\right) \go_{--}   + {1\over 4r} \tilde{g}^{(4)}_{--}    }}
In particular, this identifies $\go_{\mu\nu}$ as the conformal boundary metric appearing in \ad. 
On the other hand, $ \tilde{g}^{(4)}_{\mu\nu}  $ is not quite the same as   $\gf_{\mu\nu}$
appearing in \ad.  The discrepancy arises from the contribution of $1/r$ terms appearing in the
$\cdots$ terms in \dl.    It is clear  that  $ \tilde{g}^{(4)}_{\mu\nu}  $ and  $\gf_{\mu\nu}$
differ by an amount proportional to $\go_{\mu\nu}$.   But, as noted in the discussion after \ag, we
are not keeping track of contributions to $\gf_{\mu\nu}$ that are local in $\go_{\mu\nu}$ anyway,
since these only show up in the correlators as contact terms.    From now on, we therefore
ignore the distinction, and simply write  $ \tilde{g}^{(4)}_{\mu\nu} = \gf_{\mu\nu}$. 

The small $r$ asymptotics are
\eqn\dl{\eqalign{ p^2 \ll r \ll 1  \quad\quad\quad M_1(r) & \sim 4br \go_{++}  +{c_V \over 2b} \gf_{++}  \cr 
 N_1(r) & \sim 4br \go_{--}  +{c_V \over 2b} \gf_{--}   }}

\subsec{Matching}

Matching the expansions \de\ and \dl\ in the overlap region $  p^2 \ll r \ll 1$ gives,
\eqn\dm{\eqalign{ s_{++} & = 4b \go_{++} \cr 
 s_{--} & = 4b \go_{--} \cr  
 t_{++} & = {c_V \over 2b}  \gf_{++} \cr
 t_{-- } & = {c_V \over 2b}  \gf_{--} \cr
t_{+-} & = 0}}
along with the relations \df, which can now be written as 
\eqn\dn{\eqalign{ \gf_{++} & = {b\over 3 c_V} {p_+^3 \over p_-} \go_{--} + {b\over 3 c_V}p_+ p_- \go_{++} \cr
\gf_{--} & = {b\over 3 c_V} {p_-^3 \over p_+} \go_{++} + {b\over 3 c_V}p_+ p_- \go_{--}  }}
 We note that the last term in each line is analytic in momentum, and hence local in position space,
and thus only contributes to contact terms in correlators. 

\subsec{Correlation functions: IR behavior}

In analogy to what was done in  \bx\ for the current, we work in terms of the two-dimensional 
stress tensor defined as
\eqn\do{ \cTh_{\pm \pm} = {1\over 4} c_V V_2 \cT_{\pm\pm }}
with a corresponding relation for the expectation value of this operator.
Using \ag, \ao, and \dn, we obtain
\eqn\dpz{\eqalign{ \Th_{++} &= {c\over 24\pi}  {p_+^3 \over p_-} \go_{--} +{\rm local} \cr
 \Th_{--} &= {c\over 24\pi}  {p_-^3 \over p_+} \go_{++} +{\rm local} \cr
\Th_{+-} &= 0 +{\rm local} }}
We can now use \da\ to read off the momentum space correlators (note the factors of $2$
appearing when lowering indices using $\go_{+-}={1\over 2}$)
\eqn\dq{\eqalign{  \langle \cTh_{++}(p) \cTh_{++}(-p)\rangle & =  {c\over 48 \pi} {p_+^3 \over p_-}   \cr
 \langle \cTh_{--}(p) \cTh_{-- }(-p)\rangle & =  {c\over 48 \pi} {p_-^3 \over p_+}  }}
with all other correlators vanishing (up to contact terms).   Fourier transforming to position space gives
\eqn\dr{\eqalign{  \langle \cTh_{++}(x) \cTh_{++}(0)\rangle & = {c\over 8\pi^2}{1\over (x^+)^4} \cr
 \langle \cTh_{--}(x) \cTh_{--}(0)\rangle & = {c\over 8\pi^2}{1\over (x^-)^4} }}
These are the standard formulas for the correlation functions of the stress tensor in a D=1+1 CFT,
with $c$ being the central charge.  In particular, this demonstrates that the central charge appearing
in the IR CFT matches the Brown-Henneaux central charge defined in the near horizon AdS$_3$ region.

\subsec{Correlation functions: UV behavior}

The UV behavior of the stress tensor correlators may be computed by carrying out    
perturbation theory around pure AdS$_5$. The calculations are similar to those 
of the current correlators, and will not be given in detail here. The metric 
fluctuations now involve the Bessel function $K_2 (2p/\sqrt{r})$. From its $r \to \infty$ 
behavior, we read off the non-local contributions,
\eqn\zzd{ 4 \pi G_5 T_{\pm \pm }  =-{1\over 12}  p_\pm^4 \ln (p^2)   \go_{\mp \mp } }
The corresponding leading UV position space correlators are given by, 
\eqn\zze{
\langle \cT _{++} (x) \cT _{++} (0) \rangle = { 1 \over 4 \pi^2 G_5 (x^+)^4(x^+x^-)}}
Comparing the signs of the position space correlators of \dr\ in the IR, and 
of \zze\ in the UV, we find agreement, as is expected by unitarity in the IR
sector and the full UV theory. RG flow does not reverse this sign.

\subsec{Sub-leading contributions in the stress tensor correlators}

The stress tensor correlators exhibit an infinite series of sub-leading terms in fractional 
powers of momenta, just as the current correlators did. Their origin can be traced back
to the fluctuation modes in the field $V$, which change the size of the internal $x^{1,2}$-space.
In turn, these modes feed back into the other components of the metric. For simplicity,
we shall focus here on the correlator of the stress tensor component $\cT_V$ conjugate to $V$,
which is defined by $\cT _{ij} = \delta _{ij} \cT_V$ for $i,j=1,2$. The near region solution
for the fluctuations $h_{ij}$ is given by,
\eqn\fa{h_{ij} (r) = { v_0 \delta _{ij}  \over \sqrt{r} } \, { 2 \sin ( 2 \pi \sigma) \over \pi} 
K_{2 \sigma -1} \left ( \sqrt{ p^2 \over b^3 r} \right )}
where $v_0$ is constant.
This formula bears strong resemblance to \bk\ for the current correlators, but with 
the index $2k$ replaced by $2 \sigma -1$, where $\sigma$ was given in \am,
and the parameter $\zeta = \zeta (k)$ of \bk\ replaced by $\zeta _V$.
In the overlap region, where $p^2 \ll r \ll 1$, the behavior of the fluctuations $h_{ij}$ simplifies,
\eqn\fb{ 
h_{ij}(r) \sim v_+ r^\sigma + v_- r^{-\sigma -1}}
where we have the following formula for the ratio,
\eqn\fc{{ v_- \over v_+} = \zeta _V p^{4 \sigma +2} 
\qquad \qquad 
\zeta _V = - { \Gamma (-2\sigma ) \over \Gamma (2 + 2 \sigma) (4b^3)^{2 \sigma +1}} }
Extracting the correlator, we find the following structure,
\eqn\fd{
\langle \cT_V (p) \cT_V (-p) \rangle \sim { \zeta _V p^{4 \sigma +2} \over 1 - \zeta _V p^{4 \sigma +2} }}
This corresponding geometric series expansion may again be inferred from the presence
of a double trace interaction $\zeta _V \p_+ \cO_V \p_- \cO_V$, this time of a scalar operator $\cO_V$
of dimension $(\sigma +1/2, \sigma +1/2)$, whose two-point function is given by,
\eqn\fe{ 
\langle \cO_V (p) \cO_V (-p) \rangle = p^{4 \sigma}}
We note that the (total) dimension $4 \sigma+4$ of $\p_+ \cO_V \p_- \cO_V$ makes this perturbation 
irrelevant in the IR. The total dimension $2 \sigma+1$ of the operator $\cO_V$ itself is above the IR
Breitenloner-Freedman bound of dimension $D/2=1$. 

We close by noting that here, just as with the 
sub-leading contributions to the current correlators, there is also an infinite series of further sub-leading 
corrections in the form of integer powers of $p^2$ due to the fact that we have neglected corrections
of the type $p^2/r$ in the far region.

\subsec{Emergence of IR Virasoro algebra}

The preceding computation shows that the correlation functions of the stress tensor defined at the
AdS$_5$ boundary are consistent with the existence of an IR D=1+1 CFT corresponding to the
near horizon AdS$_3$ factor.     Associated with such a CFT is a Virasoro algebra, and it is
well known how this arises in AdS$_3$ \BrownNW.  Namely, one considers coordinate transformations that
preserve the asymptotic AdS$_3$ boundary conditions.   Such coordinate transformations include 
those that act as conformal transformations of the AdS$_3$ boundary coordinates (we will refer
to these as Brown-Henneaux coordinate transformations), and the transformation
law for the boundary stress tensor under these establishes the existence of a Virasoro algebra \BalasubramanianRE.

It is interesting to consider how the story is modified when we embed the near horizon AdS$_3$ factor in
an asymptotically AdS$_5$ geometry.   In particular, the standard boundary conditions defining
an asymptotically AdS$_5$ geometry are incompatible with an infinite dimensional group of 
coordinate transformations acting on the boundary coordinates, but naively it would seem that these are
required for the existence of the infinite dimensional  Virasoro algebra.    This puzzle can be addressed concretely using
the interpolating geometry at hand.  As we will see, the resolution is that the pure coordinate
transformations appearing in the AdS$_3$ region extend to physical modes (i.e. modes that cannot
be undone by a coordinate transformation) in the AdS$_5$ region.   The behavior of the 
AdS$_5$ stress tensor under the inclusion of these modes is what gives rise to the Virasoro algebra. 

With this motivation, we now construct a perturbed asymptotically AdS$_5$ solution whose near
geometry contains an AdS$_3$ factor perturbed by a Brown-Henneaux coordinate transformation. 
It is easiest to start in the AdS$_3$ region and then extend the solution outwards to the AdS$_5$ region. 
In our coordinates, the metric of the AdS$_3$ factor is
\eqn\ea{ ds^2 = {dr^2 \over 4 b^2 r^2} + 4br dx^+ dx^- }
Neither the ${\rm  R}^2$ factor nor the gauge field will play any role in what follows, and so we suppress them.

Now consider the following infinitesimal Brown-Henneaux coordinate transformation
\eqn\eb{\eqalign{  r & \rt r + \eps^r(r) e^{ip_+ x^+ } \cr
  x^+ & \rt x^+  + \eps^+ e^{ip_+ x^+} \cr x^-  & \rt x^- + \eps^-(r) e^{ip_+ x^+ }   }}
where $\eps^+$ is a constant and 
\eqn\ec{\eqalign{ \eps^r(r) & = -i p_+ \eps^+ r \cr
\eps^-(r) & = {p_+^2 \over 8b^3 r} \eps^+ }}
To first order in $\eps^+$ the metric becomes
\eqn\ed{ ds^2 =  {dr^2 \over 4 b^2 r^2} + 4br dx^+ dx^- + {i \over 2b^2} p_+^3 \eps^+e^{ip_+ x^+} (dx^+)^2}
This is a Brown-Henneaux coordinate transformation acting as a reparameterization of $x^+$; there
is also the obvious analogous transformation acting on $x^-$.    In terms of the 
asymptotic data written in \de\ this corresponds to 
\eqn\ee{ t_{++} =  {i \over 2b^2} p_+^3 \eps^+ }
and with $s_{++}=s_{--}=t_{--}=t_{+-}= p_-=0$. 

We can now use  formulas \dl\ and \dm\ to construct the full  asymptotically AdS$_5$ solution with
this near horizon behavior.  We simply take
\eqn\ef{ M_1 = -c_V \gf_{++} L_0^c(r)}
with
\eqn\eg{  \gf_{++} = {2b \over c_V} t_{++} }
As advertised above, the perturbation mode \ef\ is physical, and cannot be undone by a coordinate
transformation. 
Equation \eg\ shows that, up to a proportionality constant, the stress tensor measured at the
AdS$_5$ boundary is equal to the stress tensor measured at the AdS$_3$ boundary.  The
Schwarzian derivative transformation law obeyed by the AdS$_3$ stress tensor will thus be
transferred to the AdS$_5$ stress tensor.   We can also see that the proportionality constant
is precisely such that it gives the correct central charge in the Schwarzian derivative.  In particular, we
compute 
\eqn\eh{ \Th_{++} = {c\over 24 \pi} i p_+^3 \eps^+  }
or in position space,
\eqn\eh{ \Th_{++} = {c\over 24 \pi} \p_+^3  \eps^+  }
which is the Schwarzian derivative term with the correct normalization.    Of course, there is also
an ordinary tensor transformation part which is absent here since we're starting from a solution 
with vanishing stress tensor.  Finally, interchanging $\pm$ indices in this computation yields
the analogous transformation law for $\Th_{--}$.    To summarize, the AdS$_5$ stress tensor
inherits the properties of the near horizon AdS$_3$ stress tensor, yielding a pair of Virasoro algebras
with the expected central charges.  

\newsec{Conclusion}

Our objective  was to study how properties of the IR CFT dual to the near horizon
AdS$_3 \times R^2$ geometry could be extracted from correlation functions computed at
the AdS$_5$ boundary.   We found a consistent picture, after taking into account various subtleties.
In particular, to recover the IR Virasoro algebras associated with AdS$_3$ we had to take
into account the fact  that Virasoso generators do not implement pure coordinate transformations,
as they do in the asymptotically AdS$_3$ context.   And for the current correlators we had to take into
account the role played by double trace interactions. 
A pleasing aspect of these computations was that they could be carried out almost entirely analytically,
due to the simplifications occurring in the low energy limit.

In the example studied here, the IR geometry is a familiar one, since it contains an AdS factor.  
But it is worth noting that the same strategy of deducing the properties of the IR theory from the
correlators evaluated at the UV boundary can be employed in cases where the near horizon 
geometry is not so familiar, and hence where the IR CFT is not known, if it exists at all.  An example
of this occurs in our setup when we turn on a nonzero charge density.  In that case, it was shown in 
\DHokerIJ\ that the near horizon geometry contains a null warped AdS$_3$ factor.   Furthermore,
this system undergoes a quantum phase transition at a critical value of the dimensional ratio of the  
charge density to magnetic field.  These facts make it especially interesting to consider the computation
of correlation functions in these charged magnetic RG flow geometries.  
Results of such computations will be presented separately \toappear. 

\vskip .3in 

\noindent 
{ \bf Acknowledgments}

\vskip .3cm

 This work was supported in part by NSF grant PHY-07-57702 and by DARPA.

\appendix{A}{Minkowskian, Euclidean, and analytic continuation}

We have chosen to define the Einstein-Maxwell-Chern-Simons action, the current, and the stress 
tensor with respect to the Minkowski signature. The evaluation of correlators is, 
however, more cleanly  carried out with respect to the Euclidean signature. Via analytic continuation
and suitable $i\epsilon$ prescriptions we can then obtain the various Minkowski signature correlators (time ordered, retarded, advanced, etc.).   
In this appendix we provide formulas needed to effect the analytic continuation to obtain the 
Minkowski signature time ordered correlator from the Euclidean version. 

Minkowski time $t_M$ is related to Euclidean time $t_E$ by,
\eqn\apk{t_M = - i t_E} 
The analytic continuation of the variational formula \af, which defines the 
boundary current and stress tensor in terms of the Minkowski signature 
action $S=S_M$, is then given by,
\eqn\apl{i \delta S_M = \int dt_E d^3x \sqrt{\go} \left ( {1\over 2}T^{\mu\nu} \delta \go_{\mu\nu} 
+ J^\mu \delta \Ao_\mu \right)}
Note that, with our conventions for the boundary asymptotic metric $ r dx^+ dx^-$, 
the quantity $g^{(0)} = - \det (g^{(0)})$ remains positive through this analytic continuation.
Also,  it is conventional to define the Euclidean signature action $S_E$ by $-S_E=iS_M$,
but this quantity will not be needed here. As a result, the proper analytic continuations
of formulas \ba\ and \da\ are as follows,\foot{Note that the Fourier transform integral of the 
$y$-integration cancels out by the customary momentum conservation $\delta$-function.}
\eqn\apm{\eqalign{
J^\mu(x) & =   \int\! d^4 y  \sqrt{\go} \langle \cJ^\mu(x) \cJ^\nu(y) \rangle \Ao_\nu(y) \cr
T^{\mu\nu}(x) & ={1\over 2} \int\! d^4y \sqrt{\go} 
\langle \cT^{\mu\nu}(x) \cT^{\alpha\beta}(y)\rangle \go_{\alpha\beta}(y)}}
The space-time coordinates $x,y$, and the correlators, refer to Euclidean signature.
In particular, the Fourier transforms with respect to Euclidean momenta of the current
and stress tensor are given as follows,
\eqn\apm{\eqalign{
J^\mu(p) & =  \langle \cJ^\mu(p) \cJ^\nu(-p) \rangle \Ao_\nu(p) \cr
T^{\mu\nu}(p) & ={1\over 2} 
\langle \cT^{\mu\nu}(p) \cT^{\alpha\beta}(-p)\rangle \go_{\alpha\beta}(p)}}
to first order in $\Ao_\nu (p)$ and $\go_{\alpha\beta} (p)$. It is these formulas
that were used to extract the current correlators of \bz\ from \bv,  \zg\ from \ze,
and the stress tensor correlators \dq\ from \dpz, all expressed
in terms of Euclidean momenta.

\appendix{B} {Fourier transforms}

We need the Fourier transforms of various correlators from momentum space to 
position space. Let us first note a few basic conventions.
The $x^\pm$ part of the metric is written as
\eqn\zh{ ds^2 =  dx^+ dx^-}
We can view this metric either on Minkowski space for real coordinates $x^\pm = x\pm t$,
or on Euclidean space for complex coordinates $x^\pm = \sigma^1 \pm i\sigma^2$, with real $\sigma^{1,2}$.
We shall also use the notations, $p x = p_+ x^+ + p_- x^-$ and $p^2 = p_+ p_-$.
Various useful Fourier transforms in Eulcidean signature may be deduced from 
the following basic family of  integrals,
\eqn\apa{
\int { d^2 p \over (2\pi)^2} e^{i p x} \left ( p^2 \right )^a  = { 2^{2a} \, \Gamma (1+a)
\over \pi \, \Gamma (-a)  \, (x^+ x^-)^{1+a}}} 
This formula may be derived by changing variables to polar coordinates, 
integrating out the angular variable to produce a Bessel function $J_0$,
and carrying out the radial integration in terms of tabulated integrals.
Successive differentiation in $x^+$ gives,
\eqn\zm{  \int\!{d^2p \over (2\pi)^2 } e^{ip \cdot x}  (p^2)^a p_+^{2n} 
= {(-)^n \, 2^{2a} \, \Gamma( a+1+2n)  \over \pi \, \Gamma(-a) \, (x^+)^{2n} \, (x^+ x^-) ^{a+1}} }
Evaluating this expression at $a=-1$, and at $a=0$ gives, 
\eqn\apb{\eqalign{
\int { d^2 p \over (2\pi)^2} e^{i p x} { p_+^{2n} \over p^2}  & =  { (-)^n \Gamma (2n) \over 4 \pi (x^+)^{2n}} \cr
\int { d^2 p \over (2\pi)^2}  e^{i p x} p_+^{2n} \ln( p^2)  & =   { (-)^{n+1} \Gamma (2n+1) \over  \pi (x^+)^{2n} (x^+ x^-)}}}

\newsec{Appendix C: Positivity of the current algebra level in unitary theories}

Take the simple example of a charged scalar field $\phi$ in the presence of a U(1)-gauge field $A_\mu$
in Euclidean 2-dimensional space, with Euclidean action,
\eqn\ma{
S= \int d^2 x \, \left | \p_\mu \phi - i A_\mu \phi \right |^2 }
Following our standard definition, $\delta S = - \int d^2 x \, J^\mu \delta A_\mu$ the current is
found to be, 
\eqn\mb{
J_\mu = -i \left ( \phi^* \p_\mu \phi -  \phi  \p _\mu \phi ^* \right ) }
Decomposing $\phi = \phi _0 e^{i \theta}$, the free $\phi$ action reduces to 
\eqn\mc{
S = \int d^2 x \, \left ( 4 \p_+ \phi_0 \p_- \phi_0 + 4\phi _0 ^2 \p _+ \theta \p_- \theta \right ) }
while the current reduces to $J_\mu = 2 \phi _0 ^2 \p_\mu \theta$.
We now set the $\phi_0$ field to a constant,  and evaluate the $\theta$ two-point function,
\eqn\me{
\langle \theta (x) \theta (y) \rangle = - {1 \over 8 \pi \phi_0^2} \ln |x-y|^2}
The current two-point function is then readily evaluated, and we find, 
\eqn\mf{
\langle J_+ (x) J_+ (y) \rangle = - 4 \phi _0^4 \p_+ ^x \p_+ ^y \ln |x-y|^2 = - { \phi _0^2  \over 2\pi (x^+-y^+)^2} }
This formula provides the canonical normalization of the level for a free boson.

\listrefs
\end